\documentclass[a4paper,11pt]{article}
\pdfoutput=1 % if your are submitting a pdflatex (i.e. if you have
\usepackage{jcappub} % for details on the use of the package, please
\usepackage[T1]{fontenc} % if needed

\usepackage{graphicx}
\usepackage{subcaption}

\title{\boldmath Rotation Curve of the Milky Way }

\author[a]{Zi Liu,\note{Corresponding author.}}

\affiliation[a]{Department of Physics and Astronomy, University College London,
Gower Street, London WC1E 6BT, United Kingdom}

\emailAdd{liuzi2324@126.com}

\abstract{We investigate the rotation curve of the Milky Way using a multi-component mass model including a stellar disk, a gaseous disk, a bulge/bar component, and a dark-matter halo. The stellar and gas contributions are calibrated using recent observational determinations of the Galactic surface-density distribution, while the dark-matter halo is modelled with standard spherical profiles. We compute the circular-velocity contributions of the different components using a combination of spherical mass reconstruction for the bulge and halo, and thin-disk Hankel-transform methods for the disk and gas components. We first fit the stellar surface-density profile to determine a fiducial bulge--disk decomposition and then use this calibration to predict the Galactic rotation curve. We find that, although the resulting stellar mass model reproduces the observed surface-density profile reasonably well, it does not provide a fully satisfactory description of the rotation-curve data, with the largest discrepancies arising in the inner Galaxy. We then consider an alternative \emph{RC-first} calibration strategy, in which the bulge and disk parameters are adjusted to improve the kinematic fit. While this significantly improves the agreement with the observed rotation curve, the corresponding stellar surface-density profile becomes inconsistent with the independently inferred baryonic distribution. Our results highlight a tension between photometric and kinematic constraints within simplified axisymmetric models and indicate that a fully consistent description of the Milky Way mass distribution likely requires a more realistic treatment of the bulge/bar structure and of baryonic systematic uncertainties.}

\begin{document}
\maketitle
\flushbottom

\section{Introduction}
\label{sec:intro}

Rotation curves (RCs) of spiral galaxies are among the most direct dynamical
indications for the presence of non-luminous matter: beyond the region where
the baryonic mass is concentrated, the observed azimuthal speeds remain high,
implying an extended gravitating component customarily described as a dark matter (DM) halo. In the Milky Way, this conclusion is supported by decades of RC measurements employing a wide variety of tracers---including HI and
CO gas, stellar populations, masers, and other kinematic probes---spanning the inner bulge to the outer halo \cite{Sofue2020Galaxies}. Interpreting these data
requires not only accurate kinematics, but also a consistent baryonic mass model, because the inferred DM distribution depends sensitively on the
partition between luminous and dark components.

On the baryonic side, massive spectroscopic surveys combined with improved treatments of selection effects are reshaping our picture of the Milky Way stellar disk and bulge. Using APOGEE and Gaia, Lian et al.\ reconstructed the
Milky Way surface-brightness profile over an unprecedented radial range ($R\simeq 0$--$17\,$kpc) and found a broken profile with a remarkably flat
intermediate region, yielding a large half-light radius
$R_{50}=5.75\pm0.38\,$kpc \cite{Lian2024SizeMW}. Such non-exponential structure
has important implications for dynamical decompositions, because it modifies
the disk contribution to the circular speed relative to the canonical
single-exponential assumption. In a complementary analysis based on Gaia and
APOGEE, Lian revisited the local and global stellar mass budget and found a
substantially lower total stellar mass than commonly adopted, with the
difference largely driven by the flattening of the inner disk profile that
invalidates naive exponential extrapolations toward the Galactic center
\cite{Lian2025StellarMass}. Together, these results motivate updated, data-driven
stellar mass models when fitting RC data.

The gas component is likewise crucial, both as a mass contributor and as a
kinematic tracer. Modern HI and CO surveys enable increasingly detailed
reconstructions of the atomic and molecular gas distributions, and provide
rotation-curve constraints from the inner Galaxy to large radii
\cite{Sofue2020Galaxies}. However, in the bar-dominated region the use of gas
kinematics is complicated by strong non-circular motions. Recent
hydrodynamical simulations tailored to the Milky Way show that bar-driven
streaming can bias the terminal-velocity method, leading to significant
overestimates of the inferred circular speed (and enclosed mass) in the inner
$\sim$kpc if interpreted under axisymmetric assumptions \cite{Baba2025BarStreaming}.
This underscores that accurate RC-based mass inference must carefully account
for non-axisymmetries and tracer-dependent systematics, particularly at small
Galactocentric radii.

A major recent advance comes from \emph{Gaia}, which has transformed the
measurement of Galactic kinematics by providing precise parallaxes and proper
motions for large stellar samples. With Gaia DR3, rotation-curve determinations
based on stellar tracers can now extend to $\sim$30 kpc with percent-level
precision over a broad radial range, enabling stringent tests of Milky Way mass
models \cite{Beordo2024MNRAS}. These data open the possibility of performing
self-consistent fits in which the baryonic components are constrained by
state-of-the-art stellar and gas modeling, while the DM halo parameters are
inferred from the residual gravitational potential required by the observed
circular speed.

In this work we model the Milky Way circular velocity in the Solar
neighborhood using Gaia-based kinematic constraints together with a simple
four-component mass model comprising stellar disk, gas disk, bulge/bar, and a
DM halo. Each component is assigned a physically motivated density profile,
and the total circular speed is obtained from the combined gravitational
potential, allowing us to assess the baryon--DM decomposition implied by the
latest Gaia constraints.

The paper structure is the following:
in Sec.~\ref{sec:model} we report the details of the model, in Sec.~\ref{sec:results} we display our results and in Sec.~\ref{sec:conclusions} we draw our conclusions.

\section{Model}
\label{sec:model}

The Galactic RC traces the mean circular speed in the disk
mid-plane, $V_{\rm rot}(R)$, as a function of Galactocentric radius $R$, and is
therefore a direct probe of the enclosed gravitational potential and the
underlying mass distribution \cite{Sofue2020Galaxies}. In this work we adopt a
standard multi-component decomposition of the Milky Way mass model into
(i) a stellar disk, (ii) a gaseous disk, (iii) a bulge/bar component, and
(iv) a DM halo \cite{Sofue2020Galaxies,Beordo2024JCAP}.  In the
axisymmetric approximation, the squared circular speed is written as the sum
of the contributions generated by each component,
\begin{equation}
\label{eq:Vtot}
V_{\rm tot}^{2}(R)
= V_{\rm disk}^{2}(R)
+ V_{\rm gas}^{2}(R)
+ V_{\rm bulge}^{2}(R)
+ V_{\rm halo}^{2}(R)\, .
\end{equation}
This ``quadrature'' sum follows from the linearity of the gravitational
potential: each component contributes additively to the radial force, hence to
$V^2(R)=R\,\partial\Phi/\partial R$ \cite{BinneyTremaine}.

\subsection{Spherical components: bulge and halo}
For spherically symmetric components with density $\rho(r)$, the circular speed
depends only on the enclosed mass,
\begin{equation}
\label{eq:Vc_sph}
V_c^2(r)=\frac{G\,M(<r)}{r},
\qquad
M(<r)=4\pi\int_0^r \rho(r')\,r'^2\,dr'\, ,
\end{equation}
where $r$ is the spherical radius \cite{BinneyTremaine,Sofue2020Galaxies}.
This relation will be used for the (spherically-approximated) bulge component
and for the DM halo profiles considered below.

\subsection{Bulge/bar}
In the inner Galaxy the stellar bar induces significant non-circular motions,
and gas-based methods (e.g.\ terminal velocities) can substantially overestimate
the circular speed if interpreted with an axisymmetric model
\cite{Baba2025BarStreaming}. Since our primary goal is a transparent baseline
mass model for RC fitting, we approximate the bulge/bar contribution by an
effective spherical component, leaving a more complete non-axisymmetric
treatment to future work.

We model the bulge as an ``exponential spheroid'' in projection, with surface
density
\begin{equation}
\label{eq:Sigma_bulge}
\Sigma_{\rm bulge}(R)=\Sigma_{\rm b}\exp\!\left(-\frac{R}{r_b}\right),
\end{equation}
where $\Sigma_{\rm b}$ is the central surface density and $r_b$ is the bulge scale
length \cite{Sofue2020Galaxies}. In practice we convert the chosen bulge model
to a 3D density $\rho_{\rm b}(r)$ and compute $V_{\rm bulge}(r)$ through
Eq.~\eqref{eq:Vc_sph}.

\subsection{Dark-matter halo}
We test several common halo parametrizations, spanning cuspy and cored classes
\cite{Sofue2020Galaxies,Beordo2024JCAP}. For each profile we compute
$V_{\rm halo}(r)$ from Eq.~\eqref{eq:Vc_sph}.

\emph{NFW}
\begin{equation}
\label{eq:NFW}
\rho(r)=\frac{\rho_s}{(r/r_s)\,(1+r/r_s)^2}\, ,
\end{equation}
as motivated by $\Lambda$CDM N-body simulations \cite{Navarro1996,Navarro1997}.

\emph{Einasto}
\begin{equation}
\label{eq:Einasto}
\rho(r) = \rho_s \exp\!\left[-\frac{2}{\alpha}\left(\left(\frac{r}{r_s}\right)^{\alpha}-1\right)\right],
\end{equation}
a flexible simulation-motivated profile with a continuously varying logarithmic
slope \cite{Einasto1965,Beordo2024JCAP}.

\emph{Pseudo-isothermal (cored)}
\begin{equation}
\label{eq:PISO}
\rho(r)=\frac{\rho_0}{1+(r/r_c)^2}\, ,
\end{equation}
a standard cored phenomenology widely used in rotation-curve decompositions
\cite{Begeman1991}.

\emph{Burkert (cored)}
\begin{equation}
\label{eq:Burkert}
\rho(r)=\frac{\rho_0\,r_0^3}{(r+r_0)(r^2+r_0^2)}\, ,
\end{equation}
often employed to model shallow inner DM distributions \cite{Burkert1995}.

\emph{Moore (cuspy)}
\begin{equation}
\label{eq:Moore}
\rho(r) = \rho_s \left(\frac{r_s}{r}\right)^{1.5}\left(1+\frac{r}{r_s}\right)^{-1.5},
\end{equation}
representing an early steep-cusp fit to simulated halos \cite{Moore1999}.
(We adopt a common canonical form; alternative exponents exist in the
literature.)

\subsection{Implementation notes}
All disk integrals are evaluated numerically using Eq.~\eqref{eq:Vc_disk_hankel},
with $\Sigma(R)$ in units of $M_\odot\,{\rm pc}^{-2}$ (converted to $M_\odot\,{\rm kpc}^{-2}$)
and radii in kpc. We use
$G=4.30091\times10^{-6}\,{\rm kpc}\,{\rm km}^2\,{\rm s}^{-2}\,M_\odot^{-1}$.
The fitted parameter set consists of the normalizations and scale radii of the
disk and gas profiles, the bulge parameters, and the halo parameters
($\rho_s,r_s,\alpha$ etc.\ depending on the profile). The Gaia DR3 rotation-curve
constraints used in the fit follow the recent determinations based on large
samples of young disk tracers \cite{Beordo2024JCAP}.

\subsection{Axisymmetric thin disks: stars and gas}
The stellar and gaseous disks are not spherical, so the spherical relation
Eq.~\eqref{eq:Vc_sph} does not apply. For an axisymmetric thin disk specified
by a face-on surface density $\Sigma(R)$, the mid-plane circular speed can be
computed from the disk potential using a Hankel (Bessel) transform
representation \cite{BinneyTremaine}:
\begin{equation}
\label{eq:Vc_disk_hankel}
V_c^2(R)=2\pi G\,R
\int_0^\infty k\,J_1(kR)\,
\Bigg[\int_0^\infty J_0(kR')\,\Sigma(R')\,R'\,dR'\Bigg]\,
dk ,
\end{equation}
where $J_0$ and $J_1$ are Bessel functions of the first kind.  This expression
is numerically stable and avoids unphysical artefacts that can arise if one
attempts to ``spherically deproject'' a ring-like $\Sigma(R)$ (which may even
imply negative densities).
In this work, the disk surface density entering Eq.~\eqref{eq:Sigma_disk}  is modelled with a Gaussian profile,
\begin{equation}
\label{eq:Sigma_disk}
\Sigma(R) \equiv \Sigma_{\rm disk}(R)
= \Sigma_d\,
\exp\!\left[-\frac{(R-r_{d,1})^{2}}{2\,r_{d,2}^{2}}\right],
\end{equation}
where $\Sigma_d$ is the peak surface-density amplitude,  $r_{d,1}$ is the characteristic radius, and  $r_{d,2}$ is the radial width.

\paragraph{Stellar disk.}
We adopt an axisymmetric stellar surface-density profile $\Sigma_\star(R)$
motivated by recent Gaia+APOGEE reconstructions of the Milky Way radial
structure and stellar mass budget \cite{Lian2024SizeMW,Lian2025StellarMass}.
These studies indicate that a single exponential extrapolation is not adequate
over the full radial range and can bias the inferred inner mass distribution.
Given $\Sigma_\star(R)$, we compute $V_{\rm disk}(R)$ using
Eq.~\eqref{eq:Vc_disk_hankel}.

\paragraph{Gas disk.}
As a flexible phenomenological choice for the atomic gas, we employ a Gaussian
ring profile,
\begin{equation}
\label{eq:Sigma_HI}
\Sigma_{\rm HI}(R)=1.36\Sigma_0^{\rm HI}\exp\!\left[-\frac{(R-R_0^{\rm HI})^2}{2\sigma_{\rm HI}^2}\right],
\end{equation}
which captures an inner depression and a broad maximum in the outer disk.
For the molecular component we adopt an analogous ring-like profile peaked at a
few kpc, optionally supplemented by a compact central component when required
by the fit,
\begin{equation}
\Sigma_{{\rm H}_2}(r)
= 1.36(\Sigma_{1}\,\exp\!\left[-\left(\frac{r-r_{1}}{2\sigma}\right)^{2}\right]
+ \Sigma_{2}\,\exp\!\left(-\frac{r}{r_{2}}\right)
+ \Sigma_{3})\,,
\label{eq:Sigma_H2}
\end{equation}
The factor 1.36 approximately accounts for helium and metals when the
input profiles are given for hydrogen only \cite{Sofue2020Galaxies}. We model
the atomic and molecular components using axisymmetric radial profiles guided
by modern compilations of Milky Way gas distributions and RC studies
\cite{Sofue2020Galaxies,Beordo2024JCAP}. The corresponding circular-speed term
$V_{\rm gas}(R)$ is computed with the same thin-disk formalism,
Eq.~\eqref{eq:Vc_disk_hankel}.

\section{Results}
\label{sec:results}

\subsection{Calibrating the gas component to the surface-density profile}

The interstellar gas provides a non-negligible contribution to the Galactic rotation curve, especially at intermediate and large radii.  To include this baryonic component in the rotation-curve decomposition, we construct a gas circular-velocity term \(V_{\mathrm{gas}}(R)\) from published gas surface-density profiles. 
In practice, we treat the gas model as a fixed input to the rotation-curve calculation. We model the surface density of HI as Eq.~(\ref{eq:Sigma_HI}), and ${\rm H}_2$ Eq.~(\ref{eq:Sigma_H2}). We calibrate the surface-density data of HI and ${\rm H}_2$ from  \cite{MertschPhan2023} and \cite{MertschVittino2021} respectively, and references therein.

\begin{figure}[t]
    \centering
    \begin{subfigure}[t]{0.49\textwidth}
        \centering
        \includegraphics[width=\textwidth]{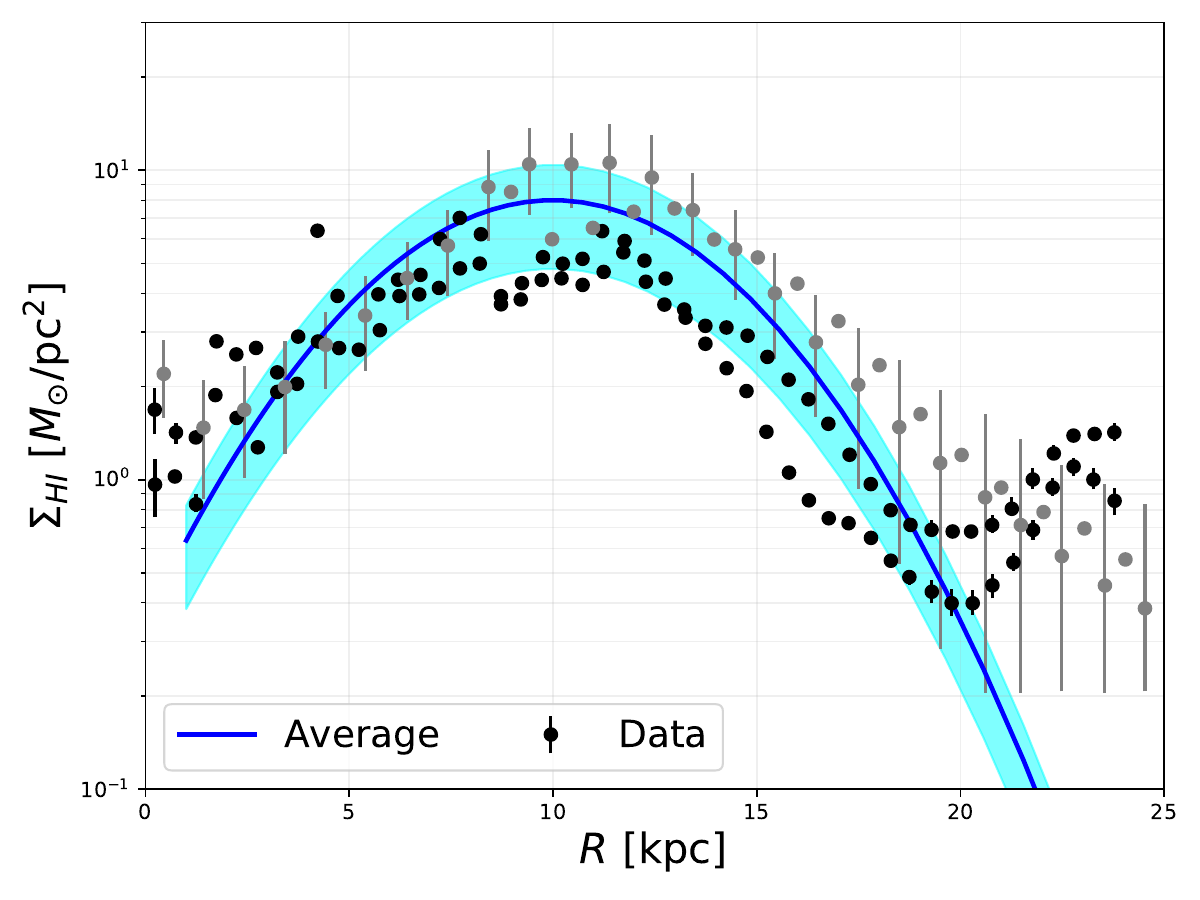}
        \caption{\textbf{Atomic gas (HI).}}
        \label{fig:sigma_hi}
    \end{subfigure}
    \hfill
    \begin{subfigure}[t]{0.49\textwidth}
        \centering
        \includegraphics[width=\textwidth]{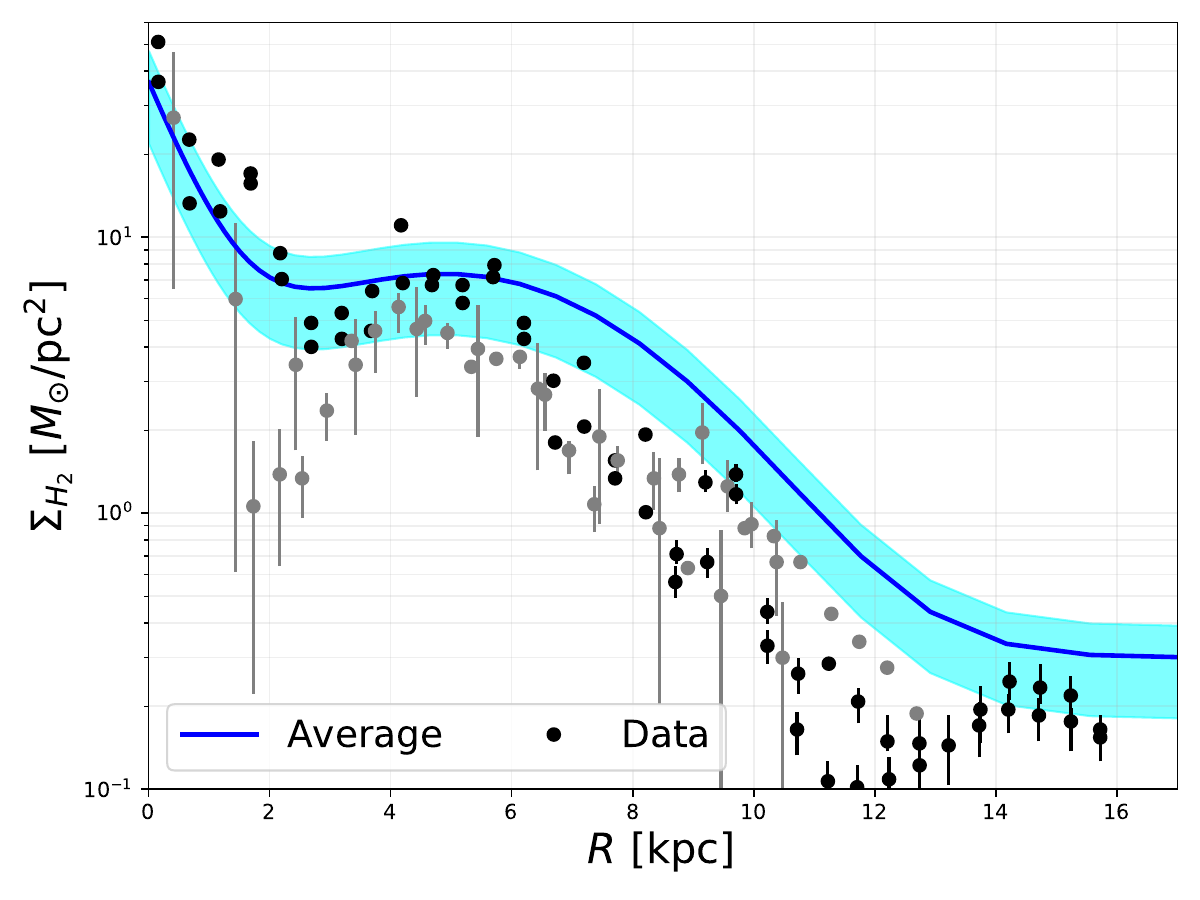} 
        \caption{\textbf{Molecular gas (H$_2$).}}
        \label{fig:sigma_h2}
    \end{subfigure}

    \caption{Adopted gas surface-density profiles used in the rotation-curve decomposition.
    The blue curve shows the adopted \emph{average} surface-density profile $\Sigma(R)$, while the shaded band provides an indicative envelope capturing systematic uncertainty in the gas distribution.
    Black points show the reference data used for comparison.
    The HI profile exhibits a broad maximum at intermediate radii, and declines towards large radii.
    The H$_2$ profile is centrally concentrated, decreases with radius more steeply than HI.
    These gas profiles are used to compute the gaseous contribution to the circular speed through $V_{\rm gas}^2(R)=V_{\rm HI}^2(R)+V_{\rm H_2}^2(R)$ and are therefore a key input for separating baryonic and DM contributions in the \emph{RC-first} calibration. The data are taken from \cite{MertschPhan2023} and \cite{MertschVittino2021} and references therein.}
    \label{fig:gas_sigma_profiles}
\end{figure}
Figure~\ref{fig:sigma_hi} and Figure~\ref{fig:sigma_h2}  show the resulting surface-density profiles obtained for the parameter choices $\Sigma_0^{\rm HI}$=8 $M_\odot\,{\rm pc}^{-2}$, $R_0^{\rm HI}$=10 kpc and $\sigma_{\rm HI}$ =4 kpc for HI (Eq.~(\ref{eq:Sigma_HI})) and $\Sigma_1$=7 $M_\odot\,{\rm pc}^{-2}$, ${r_1}$= 5 kpc, $\sigma$=2 kpc, $\Sigma_2$=35 $M_\odot\,{\rm pc}^{-2}$, ${r_2}$=0.8 kpc, $\Sigma_3$= 0.3 $M_\odot\,{\rm pc}^{-2}$ for H$_2$ (Eq.~(\ref{eq:Sigma_H2})). The parametric profiles adopted reproduce the reference surface-density data of HI and  H$_2$ well throughout the radial range, capturing both the general normalization and the principal radial trends.
We also display an uncertainty band that takes into account the scattering in the measurement of the surface-density in the literature.
%def Sigma_HI(r,Sigma0,sigma,r0):
%    return Sigma0*np.exp(-np.power((r-r0)/(np.sqrt(2)*sigma),2))

%def Sigma_HII(r,Sigma1,Sigma2,Sigma3,r1,r2,sigma):
%    part1 = Sigma1*np.exp(-np.power((r-r1)/(np.sqrt(2)*sigma),2))
%    part2 = Sigma2*np.exp(-r/r2)
%    part3 = Sigma3
%    return part1 + part2 + part3

\subsection{Calibrating disk and bulge star density}

The disk and bulge stellar components are among the main contributors to the rotational velocity.
Therefore, we calibrate these components by fitting a parametric prediction to the observed radial surface-density profile taken from \cite{Lian2025StellarMass}. The dataset is denoted by $\{(R_i, y_i,\sigma_i)\}_{i=1}^{N}$, where $R_i$ is the Galactocentric radius and $y_i$ is the measured stellar surface density at that radius. The objective is to determine the parameter vector $\boldsymbol{\theta}$ that provides the best representation of the observed profile.

The total model is constructed as the sum of a disk component and a bulge component:
\begin{equation}
y_{\mathrm{model}}(R;\boldsymbol{\theta})
=
y_{\mathrm{disk}}\!\left(R;\Sigma_d,\, r_{d,1},\, r_{d,2}\right)
+
y_{\mathrm{bulge}}\!\left(R;\Sigma_{\rm b},\, r_b\right),
\end{equation}

where $\Sigma_d$ is the disk normalization, $r_{d,1}$ and $r_{d,2}$ are disk scale parameters controlling the radial shape of the disk profile, $\Sigma_{\rm b}$ is the bulge normalization (or characteristic mass scale), and $r_b$ is the bulge scale length. The bulge term captures the centrally concentrated stellar component, while the disk term describes the more extended radial distribution.

The model parameters are determined minimizing the least-squares residual (LSR) function,
\begin{equation}
\mathrm{LSR}(\boldsymbol{\theta})
=
\sum_{i=1}^{N}
\frac{\left[
y_i - y_{\mathrm{model}}(R_i;\boldsymbol{\theta})
\right]^2}{\sigma_i^2}.
\end{equation}
This objective function measures the total squared discrepancy between the observed surface-density profile and the model prediction weighted for the data errors. Since the surface-density profile was reconstructed by digitizing published figures, point-by-point measurement uncertainties are not available in a machine-readable form. 
We therefore adopt a simple fractional error model,
\begin{equation}
\sigma_i = \epsilon\, y_i,
\label{eq:sigma_frac}
\end{equation}
with a constant $\epsilon$ applied uniformly across the dataset (we take $\epsilon = 0.2$).
This provides an approximate weighting scheme and prevents the fit from being dominated by points with large absolute surface density.

To improve numerical stability and physical interpretability, parameter bounds are imposed during the minimization. This is necessary because different combinations of
$(\Sigma_d,\, r_{d,1},\, r_{d,2},\, \Sigma_{b},\, r_b)$
may produce similarly shaped total profiles, which can lead to parameter degeneracies and unphysical solutions in an unconstrained fit. Initial parameter values are chosen from physically reasonable estimates inferred from the scale and morphology of the observed data.
The minimization is performed using the \texttt{iminuit} package. 
%In practice, the fitting procedure consists of the following steps:
%\begin{enumerate}
%    \item Define the objective function $f(\boldsymbol{\theta}) = \mathrm{LSR}(\boldsymbol{\theta})$.
%    \item Provide initial parameter estimates.
%    \item Apply parameter bounds based on physical and numerical considerations.
%    \item Run the \texttt{Migrad} algorithm to locate the minimum of the objective function.
%    \item Estimate parameter uncertainties from the minimisation output.
%\end{enumerate}
We assess the quality of the optimization by inspecting the \texttt{Minuit} convergence diagnostics (e.g.\ minimization status and validity flags) and by checking whether any best-fit parameters lie on, or very close to, the imposed parameter boundaries.
%The fitting dataset was reconstructed by digitizing the plotted data in Figure~2 of \citet{Lian2025StellarMass}. 

\begin{figure}[htbp]
    \centering
    % --- Insert the figure file here (use the exact filename in your Overleaf tree) ---
    \includegraphics[width=0.78\textwidth]{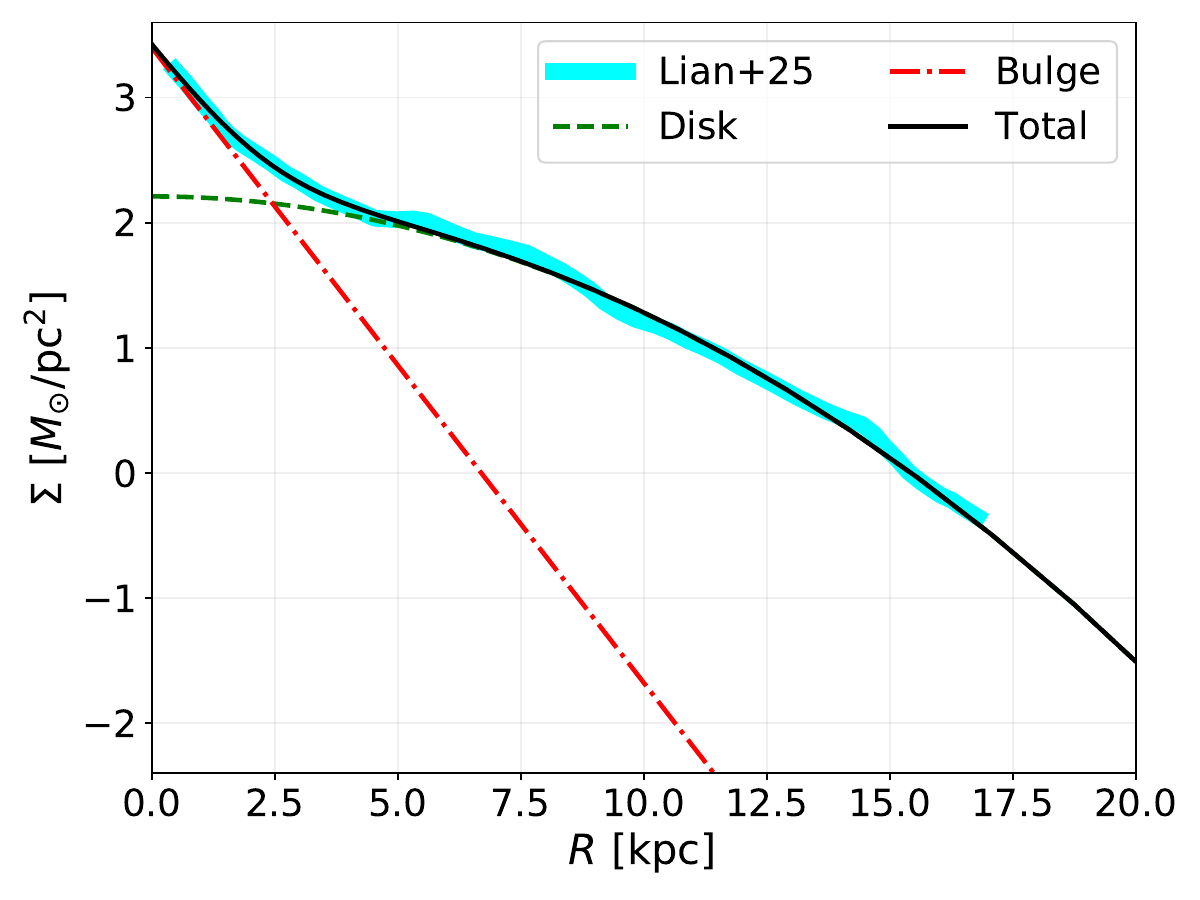}
    % ^ If your file is .png, change to: {Sigma_diskplusbulge25_lin.png}

    \caption{
    Best-fit decomposition of the stellar surface-density profile.
    Figure~\ref{fig:sigma_disk_bulge_fit} shows the stellar surface-density profile $\Sigma(R)$ extracted from \cite{Lian2025StellarMass} (cyan) together with the best-fit two-component model. The dashed and dash-dotted curves indicate the disk and bulge contributions, respectively, and the solid curve is their sum. Overall, the model reproduces the radial trend of the extracted profile over most of the fitted range, providing the calibrated bulge--disk decomposition used in the subsequent rotation-curve prediction.
    }
    \label{fig:sigma_disk_bulge_fit}
\end{figure}

Following the fitting procedure described above, we obtain the best-fitting two-component stellar surface-density model that is shown in Figure~\ref{fig:sigma_disk_bulge_fit}. We note that the best-fit model reproduces the measured stellar surface-density profile remarkably well, consistent with the results reported in \cite{Lian2025StellarMass}.
The best-fit parameters of the two-component (disk+bulge) surface-density model are summarized in Table~\ref{tab:bestfit_disk_bulge}. The uncertainties correspond to the $1\sigma$ errors returned by \texttt{iminuit}. 
We note that the typical scale radius of the bulge takes a value of about 0.7 kpc that is much smaller than the disk scale radius that is a few kpc. This is expected since the bulge star density distribution is peaked below 1 kpc where it dominates the gravitational budget.

To estimate the bulge mass implied by the fitted surface-density component, we treat the bulge as a spherically symmetric system. 
Starting from the projected bulge surface-density profile \(\Sigma_b(R)\), we reconstruct the corresponding three-dimensional density profile \(\rho_b(r)\) via Abel inversion,
\begin{equation}
\rho_b(r)
=
-\frac{1}{\pi}
\int_{r}^{\infty}
\frac{d\Sigma_{\rm bulge}/dR}{\sqrt{R^{2}-r^{2}}}\,dR,
\label{eq:abel_inversion_bulge}
\end{equation}
where \(R\) is the projected (cylindrical) radius and \(r\) is the spherical radius. 
In practice, the integral in Eq.~\eqref{eq:abel_inversion_bulge} is evaluated numerically and truncated at a sufficiently large upper limit \(R_{\rm up}\) such that the result has converged (we adopt \(R_{\rm up} = \mathrm{20}\) in this work).

The enclosed bulge mass is then computed from the reconstructed density profile as
\begin{equation}
M_b(<R)
=
4\pi
\int_{0}^{R}
r^{2}\rho_b(r)\,dr.
\label{eq:enclosed_mass_bulge}
\end{equation}
We evaluate \(M_b(<R)\) within the same radial range used for the surface-density calibration (unless stated otherwise), and report \(M_b(<R_{\max})\) as the bulge mass implied by the fitted bulge component, which is $1.12 \times10^{10}\,M_\odot$.  This is consistent with the bulge mass adopted in the literature.
For example, Lian et al.\ (2024) \cite{Lian2024SizeMW} report $M_b = (0.91 \pm 0.07)\times10^{10}\,M_\odot$, which differs from our value by $\sim 23\%$ but remains well within the combined $1\sigma$ uncertainty.

Given the best-fit disk surface-density parameters, we estimate the enclosed disk stellar mass by integrating the axisymmetric surface-density profile,
\begin{equation}
M_d(<R_{\max})
=2\pi\int_{0}^{R_{\max}}\Sigma_d(R)\,R\,dR,
\label{eq:Md_from_Sigma}
\end{equation}
where \(R_{\max}\) is chosen to match the radial range over which the surface-density profile is calibrated. In this work we adopt \(R_{\max}=\mathrm{20}~\mathrm{kpc}\). This yields a disk stellar mass of
\begin{equation}
M_d(<R_{\max}) = \,\mathrm{2.4}\times10^{10}\,M_\odot,
\end{equation}

For comparison, Lian et al.\ (2024) \cite{Lian2024SizeMW} adopt a Milky Way disk stellar mass of \(M_d=(3.6\pm0.1)\times10^{10}\,M_\odot\).

%\textcolor{blue}{Given this values of the disk and star bulge components we derive a total mass of $M_b$, $M_d$.... ADD NUMBERS. ALSO REPORT COMPARISON OF THESE VALUES WITH LITERATURE VALUES. %For example in Lian et al. 2024 ' total stellar mass of 4.81 ± 0.13 × 1010M⊙ (Methods).' and ' For the disk component, we take the measurement of 3.6±0.10 × 1010 M⊙ fr' and 'For the bulge component, we adopt the mass of 0.91±0.07 × 1010 M⊙'. From sofoue 2025 'This is two orders of magnitude smaller than the disc (∼total) mass inside R∼Rx ∼4kpc, Mdisc(≤4kpc)∼4× 1010M⊙•, and the mass of the bulge, Mbulge(≤ 4kpc) ∼ 1010M⊙•' and Lian et al 2025 ' total stellar mass of the Milky Way of 2.607±0.353(syst.)±0.085(stoch.)×1010M⊙,'

\begin{table}[htbp]
\centering
\caption{Summarize the best-fit parameters of the disk+bulge surface-density model obtained from the \texttt{iminuit} minimisation when fitting the surface-density data in \cite{Lian2025StellarMass}. The quoted uncertainties correspond to the $1\sigma$ parameter errors reported by Minuit (Hesse errors), under the adopted uncertainty model for the surface-density data.
These parameters define the fiducial bulge--disk decomposition used to compute the corresponding rotation-curve contribution of the stellar components in Figures~\ref{fig:rc_linear}--\ref{fig:rc_log}.}
\label{tab:bestfit_disk_bulge}
\begin{tabular}{lcc}
\hline
Parameter & Best-fit value ($\pm 1\sigma$) & Unit \\
\hline
$\Sigma_d$ & $79 \pm 9$     & $M_\odot\,\mathrm{pc}^{-2}$ \\
$r_{d,1}$  & $4.9 \pm 0.8$  & $\mathrm{kpc}$ \\
$r_{d,2}$  & $3.7 \pm 0.2$& $\mathrm{kpc}$ \\
$\Sigma_b$ & $2120 \pm 660$ & $M_\odot\,\mathrm{pc}^{-2}$ \\
$r_b$      & $0.7 \pm 0.1$& $\mathrm{kpc}$ \\
\hline
\multicolumn{3}{l}{Minimum LSR (Minuit \texttt{fval}) $= 3.2$} \\
\hline
\end{tabular}
\end{table}

\subsection{Fitting rotation curve data}

After obtaining the best-fit parameters from the iminuit minimization of the surface-density profile, we substitute these parameters back into the bulge and disk density models to generate the corresponding surface-density components. For each radius $R$, the circular-velocity contribution of each component is computed from the adopted dynamical prescription, and the total model rotation curve is constructed from the quadratic sum of the individual contributions,  this produces the final model curves.
\begin{figure*}[t]
    \centering
    \begin{subfigure}[t]{0.49\textwidth}
        \centering
        \includegraphics[width=\linewidth]{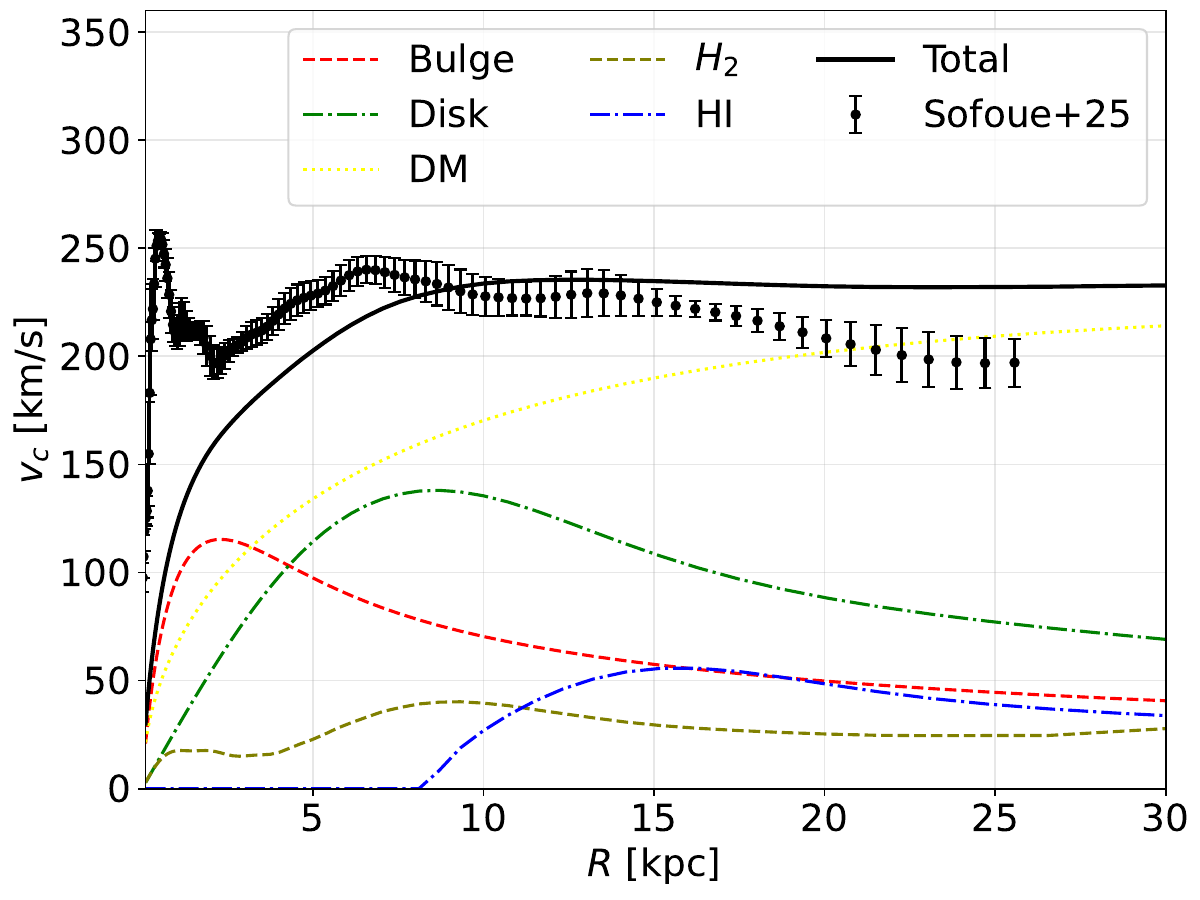}
        \caption{Linear radial scale.}
        \label{fig:rc_linear}
    \end{subfigure}
    \hfill
    \begin{subfigure}[t]{0.49\textwidth}
        \centering
        \includegraphics[width=\linewidth]{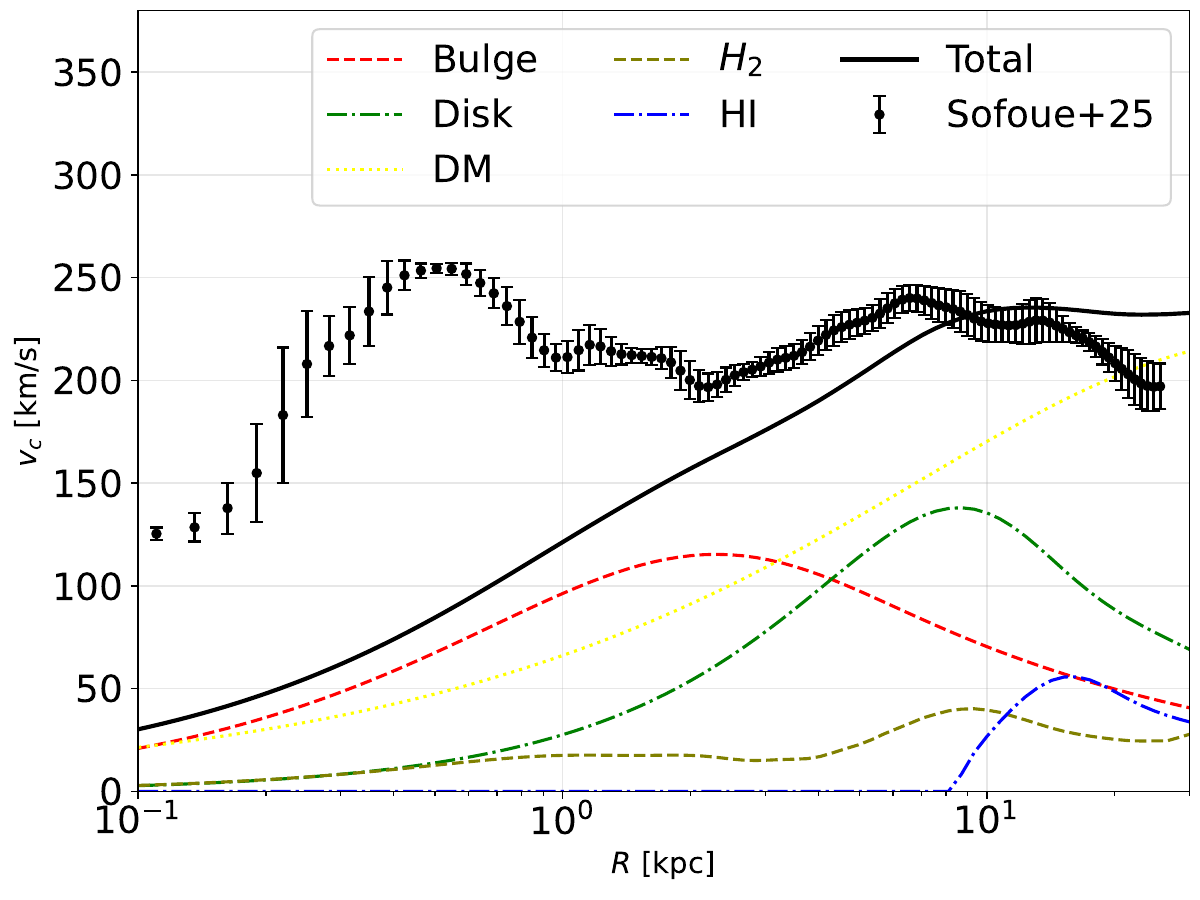}
        \caption{Logarithmic radial scale.}
        \label{fig:rc_log}
    \end{subfigure}
    \caption{Figure~\ref{fig:rc_linear} compares the observed rotation-curve data (black points with error bars) with the model prediction (solid black line), decomposed into bulge, disk, gas (HI and ${\rm H}_2$), and DM contributions. This figure highlights the disagreement at small distances between the theoretical model calibrated with the surface brightness observation of disk and bulge stars (see Figure~\ref{fig:sigma_disk_bulge_fit}) and the data. It shows also which component dominates at different radii: the bulge in the inner region, the stellar disk at intermediate radii, and the DM halo at large radii.
The decomposition provides a physical interpretation of the observed kinematics in terms of the underlying mass components. To emphasize the inner-Galaxy behaviour, Figure~\ref{fig:rc_log} shows the same rotation-curve decomposition as Figure~\ref{fig:rc_linear} but on a logarithmic radial scale. 
}

\end{figure*}

From Figure~\ref{fig:rc_linear} and \ref{fig:rc_log}, we find that  the fitted model is not fully consistent with the observed rotation-curve data. The mismatch is particularly pronounced at small radii, where the bulge contribution appears to be inadequately captured. In particular, the fitted bulge component appears too weak compared with the data, indicating that the central mass distribution is not adequately described by the present fit.

If instead we calibrate the model parameters by directly fitting RC data (\emph{RC-first} calibration strategy) rather than starting from a direct fit to the stellar surface-density profile, we first adjust the bulge--disk parameters to obtain a rotation-curve decomposition that reproduces the observed RC data as closely as possible. 
This step was carried out by iteratively varying the model parameters and inspecting the resulting model RC against the observational rotation-curve data.
Once a parameter set that provides a satisfactory RC match is identified, we then used the same parameter values to reconstruct the corresponding stellar surface-density profile, \(\Sigma(R)\), and compared it with the extracted surface-density data.

\begin{figure*}[t]
    \centering
    \begin{subfigure}[t]{0.49\textwidth}
        \centering
        \includegraphics[width=\textwidth]{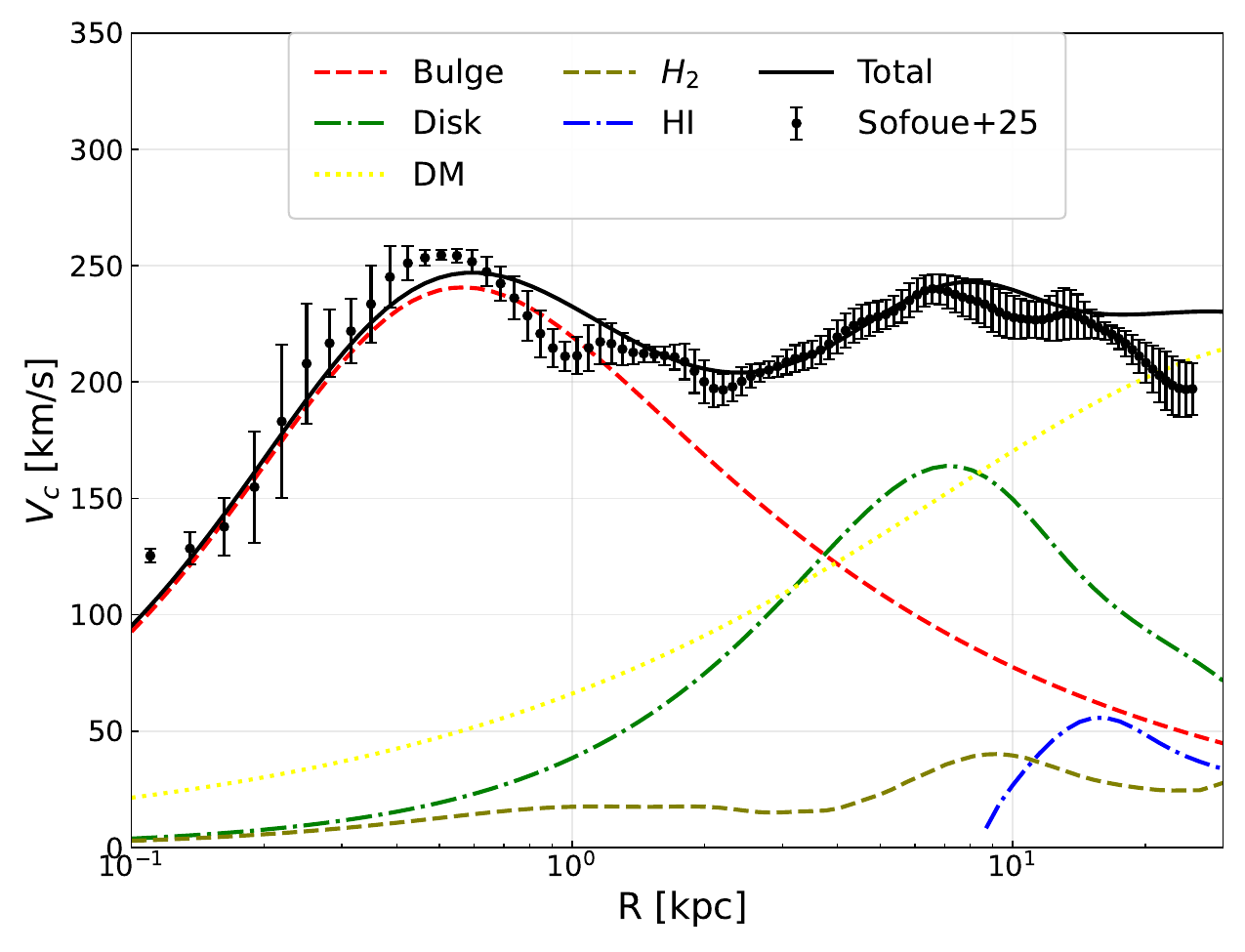} 
       \caption{Logarithmic radial scale.}
        \label{fig:rc_first_log}
    \end{subfigure}
    \hfill
    \begin{subfigure}[t]{0.49\textwidth}
        \centering
        \includegraphics[width=\textwidth]{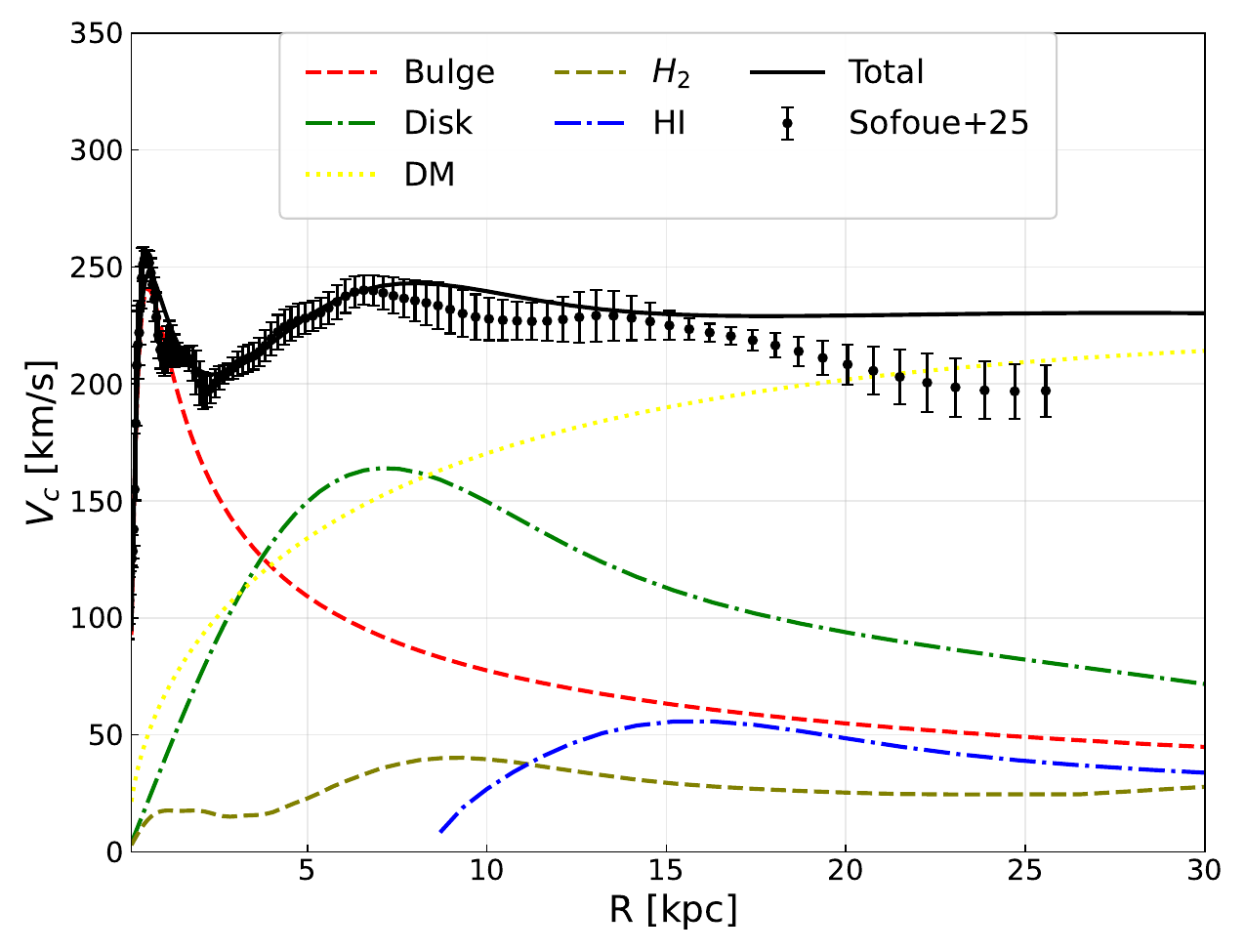} 
         \caption{Linear radial scale.}
        \label{fig:rc_first_linear}
    \end{subfigure}
  \caption{
RC-first rotation-curve decomposition of the Milky Way shown on complementary radial scales.
Black points with error bars represent the observed circular velocity $V_{\rm obs}(R)$ as a function of Galactocentric radius $R$.
The solid black curve shows the total model circular velocity,
$V_{\rm model}^2 = V_{\rm bulge}^2 + V_{\rm disk}^2 + V_{\rm gas}^2 + V_{\rm DM}^2$,
where $V_{\rm gas}^2 = V_{\rm HI}^2 + V_{\rm H_{2}}^2$.
Colored curves indicate the contributions from the bulge (red), stellar disk (green), gas components (HI/${\rm H}_2$; blue/olive), and the NFW DM halo (yellow).
Panel (a) shows the logarithmic radial scale and panel (b) the linear radial scale.
}
    
    \label{fig:rc_first_two_scales}
\end{figure*}

In the \emph{RC-first} strategy, we proceed as follows. 
First, we adopt the same parametric bulge-disk model introduced above and compute the model rotation curve \(V_{\mathrm{model}}(R)\) as the quadrature sum of the contributions of the individual components (bulge, disk, gas, and DM). 
Starting from a physically plausible initial parameter set, we vary the bulge and disk parameters (while keeping the remaining components fixed to their baseline choices) and recompute \(V_{\mathrm{model}}(R)\) after each update. 
At each iteration, we compare the model curve with the observed rotation-curve data \(V_{\mathrm{obs}}(R)\) over the full radial range and retain parameter updates that reduce the mismatch, assessed both visually and through the overall residual trend.
This iterative procedure is repeated until further parameter changes produce no appreciable improvement in the agreement with the RC data.
Second, once a parameter set that provides a satisfactory rotation-curve decomposition is obtained, we fix these parameters and use them to reconstruct the corresponding stellar surface-density profile implied by the same bulge--disk model. We then compare it directly with the extracted surface-density dataset to evaluate whether a parameter set calibrated primarily by kinematic constraints can simultaneously reproduce the photometric (surface-density) constraint.
The results are shown in Figure~\ref{fig:rc_first_log} and Figure~\ref{fig:rc_first_linear}, and the best-fit parameters are reported in Table~\ref{tab:rc_first_bestfit}.

\begin{figure}[htbp]
    \centering
    \includegraphics[width=0.82\textwidth]{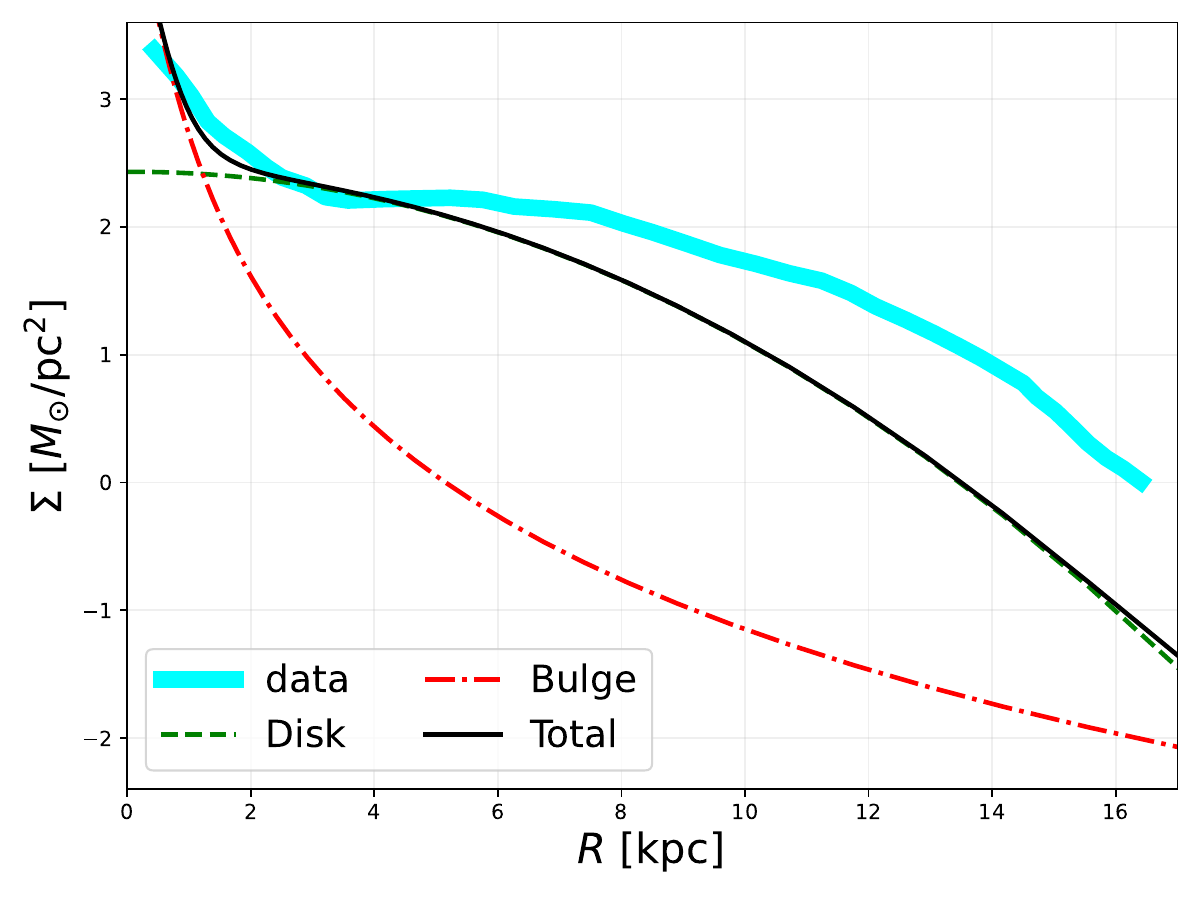}
    \caption{Stellar surface-density profile reconstructed using the RC-calibrated parameter set, shown together with the extracted \(\Sigma(R)\) data and the corresponding bulge, disk, and total model components. After tuning the bulge--disk parameters to match the rotation-curve data, we reconstruct the corresponding stellar surface-density profile $\Sigma(R)$ and compare it with the extracted surface-density dataset in Figure~\ref{fig:sigma_rc_calibrated}. Although the reconstructed profile follows the broad radial decline, the agreement with the extracted $\Sigma(R)$ data is limited, with the most pronounced discrepancy occurring in the bulge-dominated inner region. }
    \label{fig:sigma_rc_calibrated}
\end{figure}

\begin{table}[htbp]
\centering
\caption{Best-fit parameters obtained using the \emph{RC-first} calibration strategy. The parameter set is selected to reproduce the observed rotation curve and is subsequently used to reconstruct the corresponding surface-density profile for consistency checks.}
\label{tab:rc_first_bestfit}
\begin{tabular}{lcc}
\hline
Parameter & Best-fit value & Unit \\
\hline
% --- Stellar disk ---
$\Sigma_d$      & 270& $M_\odot\,\mathrm{pc}^{-2}$ \\
$r_{d,1}$       & 0.1& $\mathrm{kpc}$ \\
$r_{d,2}$       & 4.0& $\mathrm{kpc}$ \\
% --- Stellar bulge ---
$\Sigma_b$      & 0.4& $M_\odot\,\mathrm{pc}^{-2}$ \\
$r_b$           & 14000& $\mathrm{kpc}$ \\
\end{tabular}
\end{table}
However, as shown in Figure~\ref{fig:sigma_rc_calibrated}, using the RC-calibrated parameter set to reconstruct the stellar surface-density profile leads to a poor match with the extracted \(\Sigma(R)\) data. Although the model captures the overall trend, noticeable discrepancies remain between the fitted curves and the extracted surface-density profile, particularly in the disk component, indicating that the parameter set optimized for the RC does not simultaneously provide a good fit to the surface-density data.

\section{Conclusions}
\label{sec:conclusions}

In this work we have investigated the rotation curve of the Milky Way using a simple but physically motivated multi-component mass model including a stellar disk, a gaseous disk, a bulge/bar component, and a DM halo. The main goal was to assess whether recent Gaia-based kinematic constraints can be simultaneously reproduced alongside independent information on the baryonic mass distribution, in particular the stellar surface-density profile inferred from recent observational reconstructions.

We first constructed a baseline Galactic mass model in which the bulge and DM halo were treated as spherical components, while the stellar and gaseous disks were modelled as axisymmetric thin disks. The gas contribution was calibrated using parametric profiles for the H\,\textsc{i} and H$_2$ components tuned to the results of \cite{MertschPhan2023,MertschVittino2021}, while the stellar disk and bulge parameters were obtained by fitting the observed stellar surface-density profile reported in \cite{Lian2025StellarMass}. This procedure yields a reasonable decomposition of the stellar mass distribution, with bulge and disk masses broadly consistent with recent estimates, and therefore provides a plausible baryonic model from the photometric point of view.

However, once these surface-density-calibrated baryonic components are inserted into the dynamical model, the resulting rotation curve does not fully reproduce the observed Galactic kinematics. In particular, the discrepancy is most evident in the inner Galaxy, where the predicted bulge contribution appears too weak to account for the measured circular speed. This indicates that a calibration based primarily on the stellar surface-density profile does not automatically translate into an equally successful description of the rotation curve, especially in the central few kiloparsecs.

To explore this tension further, we also adopted an alternative \emph{RC-first} strategy, in which the bulge--disk parameters were adjusted directly to improve the agreement with the observed rotation curve. This approach leads to a significantly better kinematic description, especially at small radii where the bulge dominates. Nevertheless, the corresponding reconstructed stellar surface-density profile shows a poor match to the independently inferred $\Sigma(R)$ data in \cite{Lian2025StellarMass}. Therefore, the parameter set that best reproduces the kinematics is not compatible with the one preferred by the stellar surface-density constraint.

Taken together, these results point to a non-trivial tension between photometric and dynamical calibrations within the simplified axisymmetric framework adopted here. This mismatch may reflect several effects. First, the inner Milky Way is known to be strongly non-axisymmetric because of the Galactic bar, and the spherical approximation used for the bulge is likely too crude to capture the true central mass distribution. Second, systematic uncertainties in the gas and stellar profiles, including the reconstruction of the surface-density data and the assumed error model, may affect the inferred baryonic decomposition. Third, the DM halo parametrization may partly absorb mismodelling of the baryonic components, leading to degeneracies between halo and bulge/disk parameters.

Our analysis therefore highlights both the usefulness and the limitations of simple rotation-curve decompositions. On the one hand, they provide a transparent framework to separate the relative roles of bulge, disk, gas, and DM in shaping the Galactic circular velocity. On the other hand, the Milky Way appears to require a more refined treatment if one aims at a genuinely self-consistent fit to both kinematic and surface-density observables.

A natural extension of this work would be to incorporate a more realistic non-axisymmetric bulge/bar model, allow for a broader class of stellar and gas profiles, and perform a global statistical fit in which rotation-curve and surface-density data are fitted simultaneously. It would also be important to propagate systematic uncertainties associated with tracer selection, bar-induced streaming motions, and the reconstruction of the baryonic mass distribution. Such improvements will be necessary to derive robust constraints on the Milky Way DM halo and to quantify more reliably the baryon--DM decomposition implied by modern Galactic data.

Overall, this study shows that recent Gaia-era measurements provide strong leverage on the mass structure of the Milky Way, but also make clear that a satisfactory interpretation of the Galactic rotation curve cannot rely on oversimplified baryonic modelling. Achieving a consistent description of both the observed kinematics and the stellar mass distribution remains an open and important problem for future work.

%\appendix
%\section{Some title}
%Please always give a title also for appendices.

\section*{Acknowledgements}
\label{Acknowledgements}

The author gratefully acknowledges the guidance and helpful discussions provided by the project supervisor Mattia Di Mauro throughout this work. His comments and suggestions significantly improved the clarity of the analysis and interpretation. This study made use of published Galactic rotation-curve measurements and stellar surface-density reconstructions referenced throughout the text.


\bibliographystyle{apsrev4-2}
\bibliography{main}

\end{document}